\documentclass[fleqn,10pt]{wlscirep}
\title{Three-state Majority-Vote Model on Barab\'asi-Albert and Cubic Networks and the Unitary Relation for Critical Exponents}

\usepackage[utf8]{inputenc}
\usepackage[LGR,T1]{fontenc}

\author[1,2]{Andr\'e L. M. Vilela}
\author[2]{Bernardo J. Zubillaga}
\author[2,3*]{Chao Wang}
\author[2,4,5]{Minggang Wang}
\author[2,6]{Ruijin Du}
\author[2]{H. Eugene Stanley}

\affil[1]{Universidade de Pernambuco, Escola Polit\'ecnica de Pernambuco, 50720-001, Recife, Pernambuco,
  Brazil}
\affil[2]{Center for Polymer Studies and Department of Physics,
  Boston University, Boston, Massachusetts, 02215, USA}
\affil[3]{College of Economics and Management, Beijing University of Technology, Beijing, 100124, China}
\affil[4]{School of Mathematical Science, Nanjing Normal University, Nanjing 210042, Jiangsu, China}
\affil[5]{Department of Mathematics, Nanjing Normal University Taizhou College, Taizhou 225300, Jiangsu, China}
\affil[6]{Institute of Applied System Analysis, Jiangsu University, Zhenjiang, 212013, Jiangsu, China}
\affil[*]{Corresponding author: chaowanghn@vip.163.com}

\begin{abstract}

We investigate the three-state majority-vote model with noise on scale-free and regular networks. In this model, an individual selects an opinion equal to the opinion of the majority of its neighbors with probability $1 - q$ and opposite to it with probability $q$. The parameter $q$ is called the noise parameter of the model. We build a network of interactions where $z$ neighbors are selected by each added site in the system, yielding a preferential attachment network with degree distribution $k^{-\lambda}$, where $\lambda \sim 3$. In this work, $z$ is called growth parameter. Using finite-size scaling analysis, we show that the critical exponents associated with the magnetization and magnetic susceptibility add up to unity when a volumetric scaling is used, regardless of the dimension of the network of interactions. Using Monte Carlo simulations, we calculate the critical noise parameter $q_c$ as a function of $z$ for the scale-free networks and obtain the phase diagram of the model. We find that the critical noise is an increasing function of the growth parameter $z$, and we define and verify numerically the unitary relation $\upsilon$ for the critical exponents by calculating $\beta /\bar\nu$, $\gamma /\bar\nu$ and $1/\bar\nu$ for several values of the network parameter $z$. We also obtain the critical noise and the critical exponents for the two and three-state majority-vote model on cubic lattices networks where we illustrate the application of the unitary relation with a volumetric scaling.

\end{abstract}
\begin{document}

\flushbottom
\maketitle
\thispagestyle{empty}

\section*{Introduction}

\begin{figure*}[t]
  \centering
    \includegraphics[width = 0.77\linewidth]{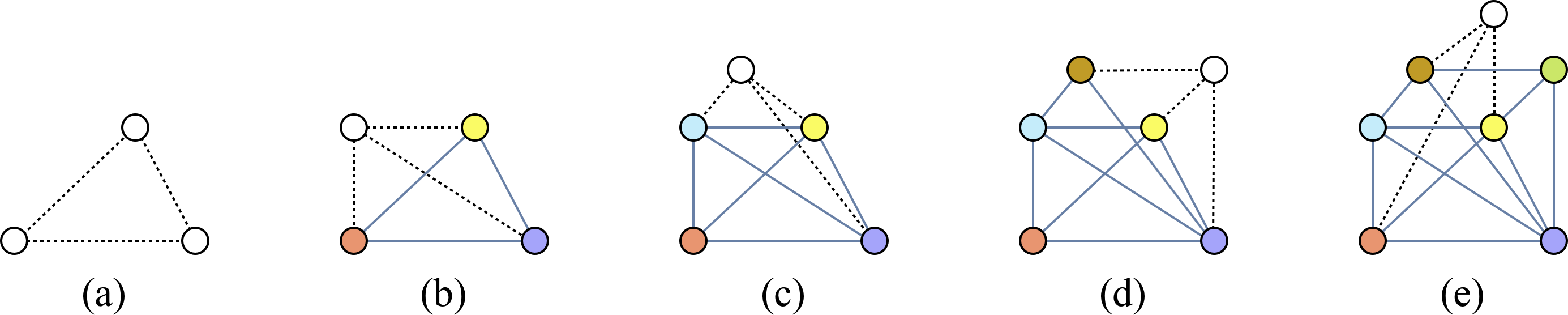}
    
\caption{The sequence shows five subsequent steps of the Barab\'asi-Albert model for $z = 3$. Empty circles mark the newly added node to the network and dashed lines represents its new links, which decides where to connect using the preferential attachment.}

  \label{barabasi-albert}
\end{figure*}

Regular networks and random graphs have been used to study and describe the topology of diverse systems investigated in condensed matter physics, but they do not represent several behaviors of real networks found in nature. \cite{Barabasi1999, Barabasi2000, Barabasi2002, Newman2003, Newman2010, Fontoura2011} Using complex networks, physicists studied a wide variety of physical systems such as the internet, the world wide web, the cellular networks, Protein-protein interaction networks, the scientific collaboration network, airline networks, economic and financial markets, among others. \cite{Faloutsos1999, Newman2001, Petre2013, Verma2014, Vina2016, Deborah2016, Gosak2018, Mario2018, Vilela2019} Many real systems are ordered in networks that present a universality of topology, showing the same architectures of assembly. One of the most investigated kinds of networks present in real-world systems are the scale-free networks or Barab\'asi-Albert networks. \cite{Barabasi1999, Barabasi2000, Barabasi2002} These networks can be built by starting with an initial number of interconnected nodes, and newly added nodes have a higher probability to attach to the more connected nodes in a mechanism known as preferential attachment. In this process, highly connected nodes acquire more links than those that have fewer connections, yielding sites with a high number of connections, or hubs of the network. The degree distribution of these networks presents a power-law decay with exponent $\lambda \sim 3$. In Fig. \ref{barabasi-albert} we illustrate the preferential attachment algorithm for a network where each newly added site connects to others with growth parameter $z = 3$.

The three-state majority-vote model with noise defined on a regular square lattice is a system of spins, where each one is allowed to be on three states only. \cite{Tome1999, Tome2002, Melo2010, Lima2012, Balankin2017} In this three-state model, each spin assumes the state of the majority of its neighboring spins with probability $(1-q)$ and the opposite state with probability $q$, which is known as the noise parameter of the model. The increase of the parameter $q$ promotes the formation of opposite opinion configurations in the model, and $q$ acts as a social temperature, disordering the opinion system. When the social interactions of the model are modeled on a regular square lattice network, it presents an order-disorder phase transition at the critical value $q_c \sim 0.118$ for the noise parameter.

In this work, we investigate the influence of a network with preferential attachments on the three-state majority-vote model with noise. We use Monte Carlo simulations and standard finite-size scaling techniques to determine the critical noise parameter $q_{c}$ and to obtain the phase diagram of the system, as well as the critical exponents for several values of the growth parameter $z$ of the networks investigated. We propose a unitary relation to verify the criticality of the system obtained by a volumetric scaling. We conjecture that this relation is universal regardless of the network of social interactions. We also perform simulations for the three-state majority-vote model in cubic networks that confirms our results and obtain the critical noise for this system and its critical exponents.

This work is organized as follows. In section II we describe the non-equilibrium three-state majority-vote model with noise, the network construction process and introduce the relevant quantities used in our simulations. Section III contains our results in complex and regular networks, along with a discussion. In section IV we present our conclusions and final remarks.

\section*{The Model}

\subsection*{The Barab\'asi-Albert Network}
The three-state majority-vote model with noise consists in a set of spin variables $\{\sigma_{i}\}$ with $i = 1, 2,..., N$, where each spin can assume one of the values $\sigma = 1, 2,$ or $3$, representing the opinion for an individual. We place the individuals in the nodes of a scale-free network with $N$ sites. In this context, we build our network from a core of $z$ fully connected nodes, where we add new nodes - one at a time - with $z$ free links which will be connected by preferential attachment to the existing nodes of the network. In other words, the probability $\Pi (k_i)$ that a link of the new node $j$ connects to node $i$ depends on the degree $k_i$ of the node $i$. Thus, for Barab\'asi-Albert networks with linear preferential attachment we write 
\begin{equation} \label{probpi}
\Pi (k_i) = \frac{k_i}{\sum_{\ell}{k_\ell}},
\end{equation}

\noindent where the summation is equal to the total number of existing links in the network. We keep adding nodes to the network until it reaches a total of $N$ sites, where a double connection to the same site is forbidden.

\begin{figure}
\centering
  \includegraphics[width=0.40\linewidth]{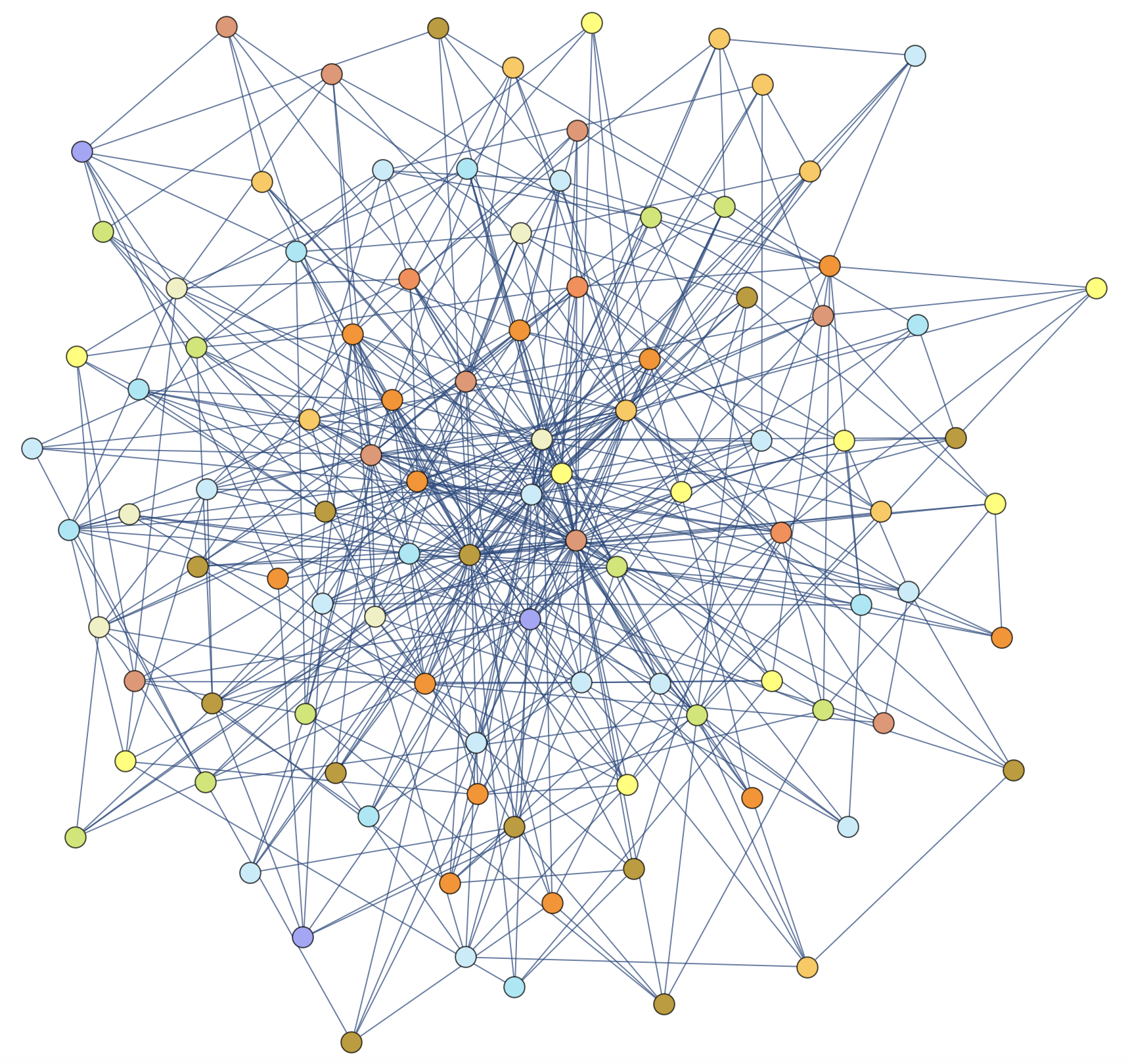}
  \includegraphics[width=0.49\linewidth]{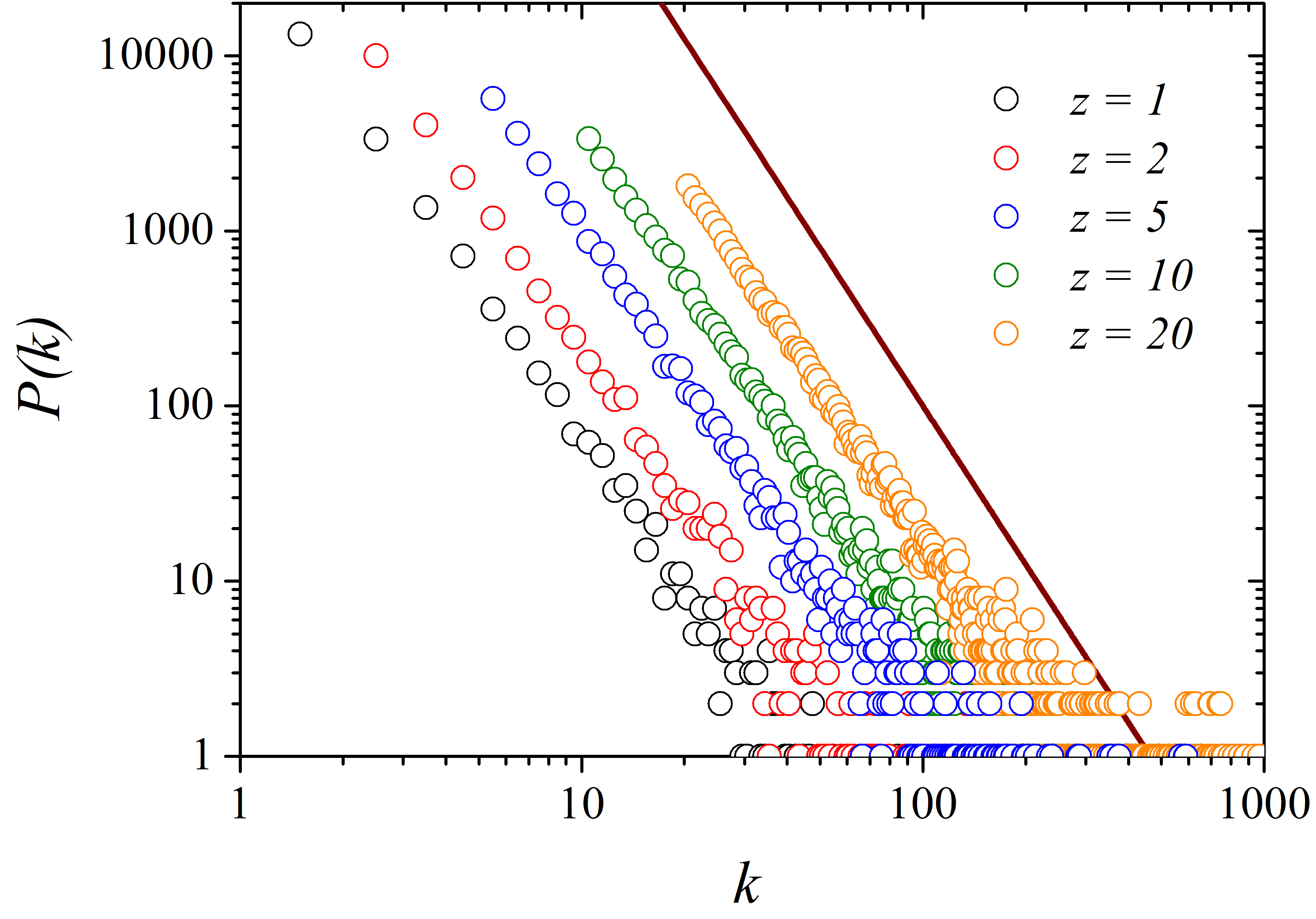}\\
  \caption{A Barab\'asi-Albert network representation for $N = 100$ nodes with $z = 5$ (left), and the degree distribution histogram $P(k)$ for a single network of size $N = 20000$ and $z = 1, 2, 5, 10$ and $20$ (right). The straight line is a guide to the eye and has slope corresponding to the network’s predicted degree exponent of decay $\lambda \sim 3$.
  }
  \label{pk}
\end{figure}

In Fig. \ref{pk} we show a Barab\'asi-Albert network representation for $N = 100$ sites with growth parameter $z = 5$ (left). Note that some nodes have a high number of connections, despite the average small value of connections, or degree, per site in the network. We also present the histogram of the degree of the nodes for networks with $N = 20000$ nodes and different values of the growth parameter $z$ (right). We verify that our scale-free networks present the characteristic power-law decay with exponent $\lambda \sim 3$, even for different values of the average degree per node $z$ as expected. \cite{Barabasi2000}

\subsection*{Dynamics and Numerical Quantities}
The dynamics of the system consists in a generalization for three states of the two-state majority-vote model. \cite{Oliveira1992, Felipe2005, Lima2006, Vilela2018} For each randomly selected spin $\sigma_{i}$ we determine the opinion of the majority of the spins that are linked to it. With probability $1 - q$ the selected spin adopts the same opinion of the majority of its neighbors, and with probability $q$ adopts one opposite opinion. In the case of a tie between all the three states, the selected spin $\sigma_{i}$ changes to any opinion with the same probability equal to $1/3$. For the case of a tie between two majority opinions, $\sigma_{i}$ assumes one of these tied opinions with probability $(1-q)/2$, and the minority with probability $q$. Finally, for the case of a single majority opinion, $\sigma_{i}$ assumes one of the two minority opinions with equal probability $q/2$, and the majority with probability $1-q$. That is, if $n_{\alpha}$ is the number of neighbors of the spin $\sigma_i$ in a given state $\alpha = 1, 2, 3$, we can write the following probabilities for $\sigma_i$ to assume the opinion $1$:
\begin{equation}\label{rules}
\begin{array}{l}
P(1 | n_1 > n_2, n_3 )  =  1 - q, \qquad P(1 | n_1 = n_2 > n_3 )  =  (1 - q)/2, \qquad
P(1 | n_1 < n_2 = n_3 )  =  q, \\
\\
P(1 | n_1, n_2 < n_3 )  =  q/2, \qquad \ \
P(1 | n_1 = n_2 = n_3 )  =  1/3.
\end{array}
\end{equation}
These transition rules present the $C_{3 \nu}$ symmetry with respect to the simultaneous change of all opinions, and the probabilities for the other states $\sigma = 2$ and $3$ are obtained by the symmetry operations of the $C_{3 \nu}$ group. The total number of individuals connected to $\sigma_i$ is $n = n_1 + n_2 + n_3$. The probability $q$ is the noise parameter of the model and all probabilities satisfy
\begin{equation} \label{prob1}
P(1 | ...) + P(2 | ...) + P(3 | ...) = 1.
\end{equation}

To investigate the critical behavior of the three-state majority-vote model, we first calculate the average opinion, defined in analogy to the three-state Potts model
\begin{equation} \label{magvector}
m = (m_1^2+m_2^2+m_3^2)^{1/2},
\end{equation}

\noindent whose normalized components are given by
\begin{equation} \label{magcomp}
m_{\alpha} = \sqrt{\frac{3}{2}} \left[ \frac{1}{N} \sum_{i=1}^{N} \delta( \alpha, \sigma_i) - \frac{1}{3} \right],
\end{equation}

\noindent where the sum is over all sites in the network of social interactions and $\delta(\alpha, \sigma_i)$ is the Kronecker delta function. In this way, to study the critical behavior of the model we consider the magnetization $M$, the magnetic susceptibility $\chi$, and the Binder's fourth-order cumulant $U$ defined by
\begin{equation} \label{mag}
M(q, z, N) = \left\langle \left\langle m \right\rangle _{t} \right\rangle _{c},
\end{equation}
\begin{equation} \label{suscep}
\chi(q, z, N) = N\left[ \langle \langle m^{2}\rangle _{t} \rangle _{c} - \langle
\langle m \rangle _{t} \rangle _{c}^{2}\right],
\end{equation}
\begin{equation} \label{cumul}
U(q, z, N) = 1 - \frac{\langle \langle m^4 \rangle _{t} \rangle _{c} }{3\langle
\langle m^2 \rangle _{t} \rangle _{c}^2},
\end{equation}
\noindent where $q$ is the noise parameter, $z$ is the growth parameter of the network, $N$ is the total number of sites, $\langle ... \rangle_{t}$ denotes time averages taken in the stationary regime and $\langle ... \rangle_{c}$ stands for configurational averages. The critical behavior of the model is investigated by performing computer simulations and using finite-size scaling analysis.

The three-state majority-vote model evolves in time according to the probability rules given by Eqs. (\ref{rules}) and eventually reaches a steady state that can be of two types. For $q = 0$ the system exhibits an ordered steady state characterized by the predominance of individuals in one of the possible opinions. Assuming that $\sigma = 1$ for all sites, one can write $m_1 = \sqrt{2/3}$, $m_2 = m_3 = -1/\sqrt{6}$ and $m = 1$, yielding $M = 1$ for $q = 0$. The upper limit for $q$ is obtained when the probability of a given spin agreeing with the majority of its neighbors is equal to the probability of it agreeing with any of the other two minority states, thus $1 - q = q/2 \Rightarrow q = 2/3$. In this case, any state $\sigma = 1, 2$ or $3$ can be found with equal probability [Eqs. (\ref{rules})], leading to $m_1, m_2, m_3 \simeq 0$ and $M = 0$ for $q = 2/3$ in the thermodynamic limit $N \to \infty$.

\section*{Discussion and Results}
\subsection*{Monte Carlo Simulations}
We perform Monte Carlo simulations on Barab\'asi-Albert networks with sizes ranging from $N = 1000$ to $20000$. For each value of the pair $q$ and $z$, we set all spins to point to one opinion, i.e., $\sigma_i = 1$ for all $i$ in the network. We next select a randomly chosen individual and update its opinion in accordance with the rules given by Eqs. (\ref{rules}). A Monte Carlo step (MCS) is accomplished after updating all $N$ spins. We skip $10^5$ MCS in the simulation to overcome transients and allow the system to reach a steady state. The time averages were estimated from the next $2 \times 10^5$ MCS and we generated at least $100$ independent random samples in order to calculate the configurational averages. Starting from different initial opinion configurations, we find that the final steady state of the system falls into becoming ordered or disordered, depending on $q$, $z$ and $N$. In the ordered phase the majority of opinions are found in one of the possible states $1, 2$ or $3$. In the disordered phase, the three opinions are equally distributed in the network of social interactions.

\begin{figure*}[t]
  \centering
    \includegraphics[width=0.9\linewidth]{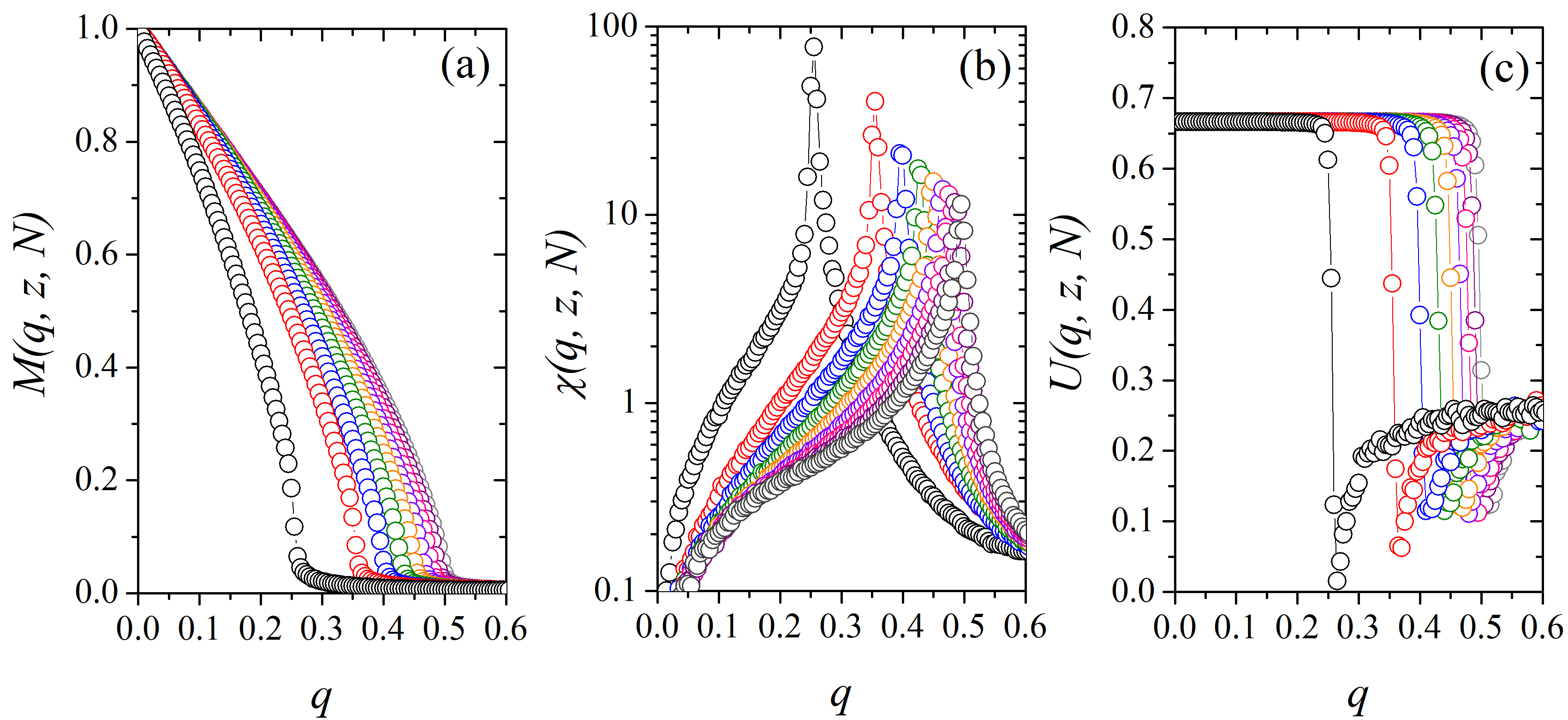}
    
\caption{(Color online) Three-state majority-vote model on Barab\'asi-Albert networks for $N = 20000$. (a) Magnetization $M(q, z, N)$, (b) susceptibility $\chi(q, z, N)$ and (c) Binder's fourth-order cumulant $U(q, z, N)$ as a function of the noise parameter $q$ for $z = 2, 3, 4, 5, ..., 10$, from left to right. Lines are just a guide to the eyes.}

  \label{overmxu}
\end{figure*}

In Fig. \ref{overmxu} we illustrate the effect of the network of interactions with preferential attachment in the consensus (order) of the system. We plot the magnetization $M(q, z, N)$, the susceptibility $\chi(q, z, N)$ and the Binder's fourth-order cumulant $U(q, z, N)$ as a function of the noise parameter $q$ for $N = 20000$ and $z = 2, 3, 4, 5, ..., 10$. We find that for small values of the noise parameter $q$ the system presents an ordered state where $M(q, z, N) >> 0$. Increasing $q$ the magnetization will continuously decrease to zero near a critical value $q_c$, denoting the second-order phase transition of the system. The magnetic susceptibility $\chi(q, z, N)$ exhibits a peak near the critical value of the noise parameter $q_c$ where the transition occurs, also denoted by the rapid decrease of the Binder's fourth-order cumulant $U(q, z, N)$. These results show that the critical noise parameter $q_c$ that drives the system to a disordered state is an increasing function of the growth parameter of the network $z$.

\begin{figure*}[t]
  \centering
    \includegraphics[width=0.9\linewidth]{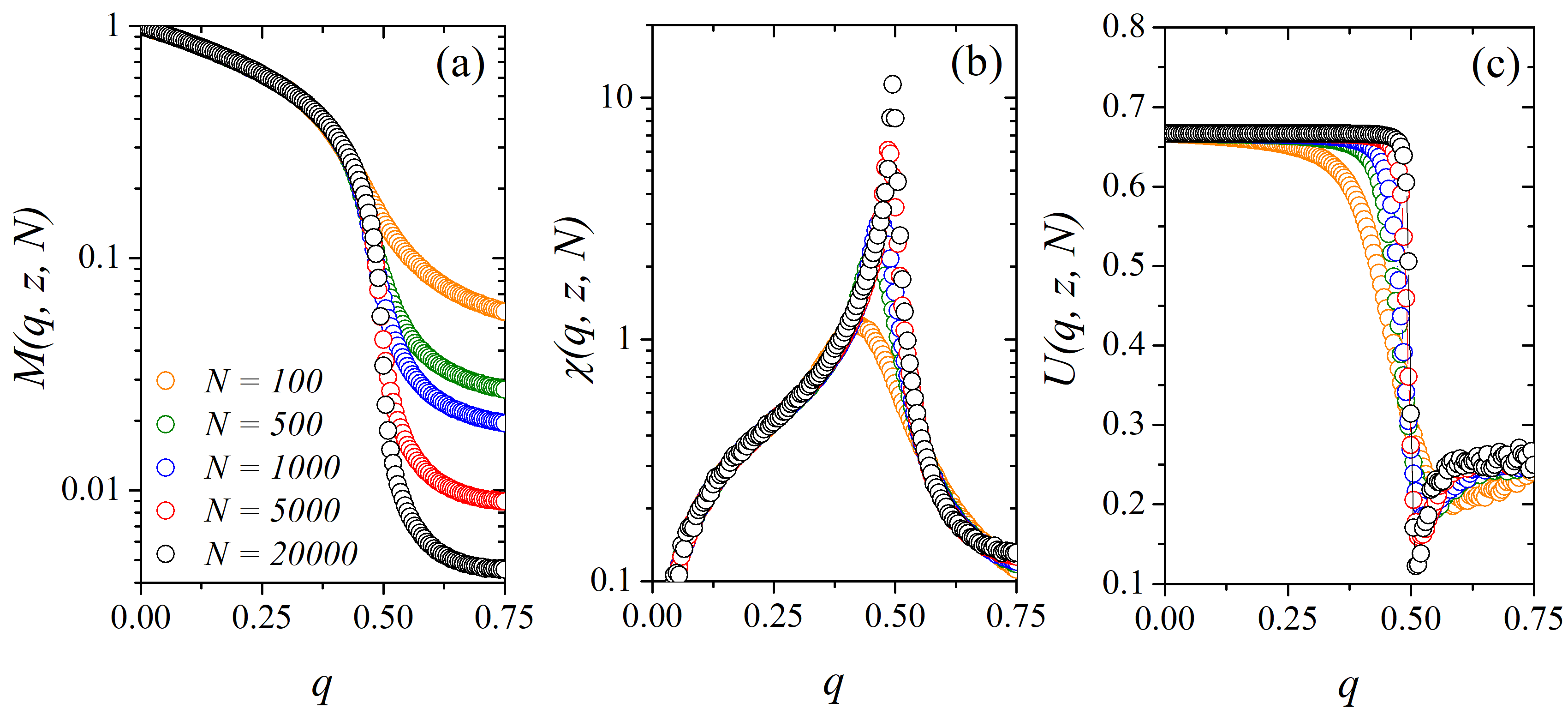}
    
\caption{(Color online) (a) Magnetization $M(q, z, N)$, (b) susceptibility $\chi(q, z, N)$ and (c) Binder's fourth-order cumulant as a function of the noise parameter $q$ for several values of the system size $N$ with $z = 10$. The critical noise for this value of the growth parameter was found to be $q_c = 0.513(1)$ and it was obtained in the curves intersection point for Binder cumulant (c).}
\label{mxuN}
\end{figure*}

Next we consider the finite-size effects on our measured quantities. In Fig. \ref{mxuN} we show the (a) magnetization $M(q, z, N)$, (b) the susceptibility $\chi(q, z, N)$ and (c) the Binder's fourth-order cumulant $U(q, z, N)$ versus the noise parameter $q$ for $z = 10$ and different system sizes $N$. Note that $M(q, z, N) \neq 0$ for high values of the noise parameter $q$ due to finite-size effects. The susceptibility $\chi(q, z, N)$ exhibits a sharper peak as we increase the system size and the position of its maximum in the horizontal axis depends on $N$. Thus, we write the pseudocritical noise as $q_c(z, N)$. In Fig. \ref{mxuN}(c) we show the Binder's fourth-order cumulant $U(q, z, N)$ as a function of $q$ for different system sizes. The critical noise parameter for the system $q_c(z)$ can be estimated as the point where the curves of $U(q, z, N)$ for different sizes $N$ intercept each other. We estimate for this set of parameters $q_c = 0.513(1)$.

\begin{figure*}[t]
  \centering
    \includegraphics[width=.414\linewidth]{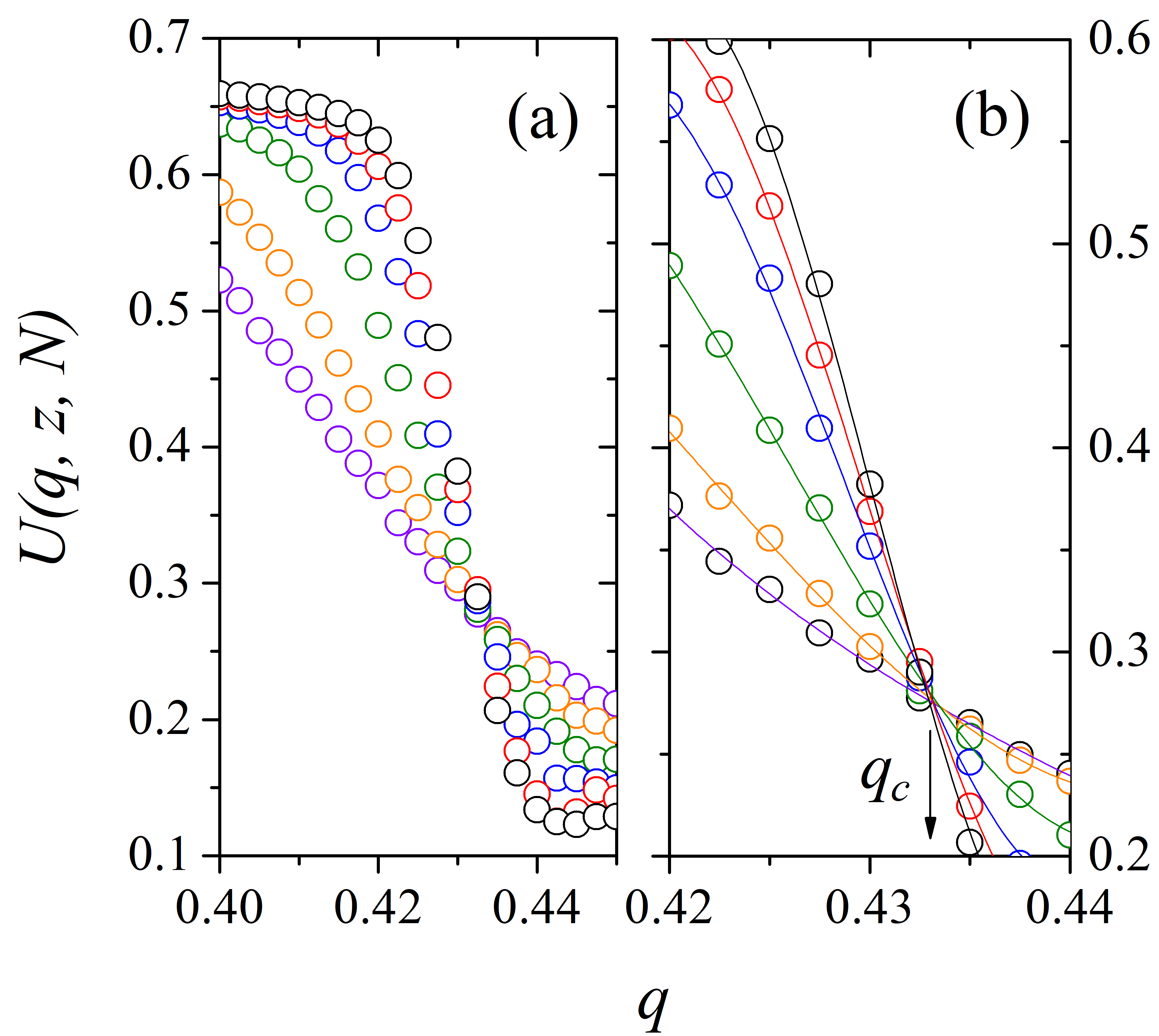}
    \includegraphics[width=.477\linewidth]{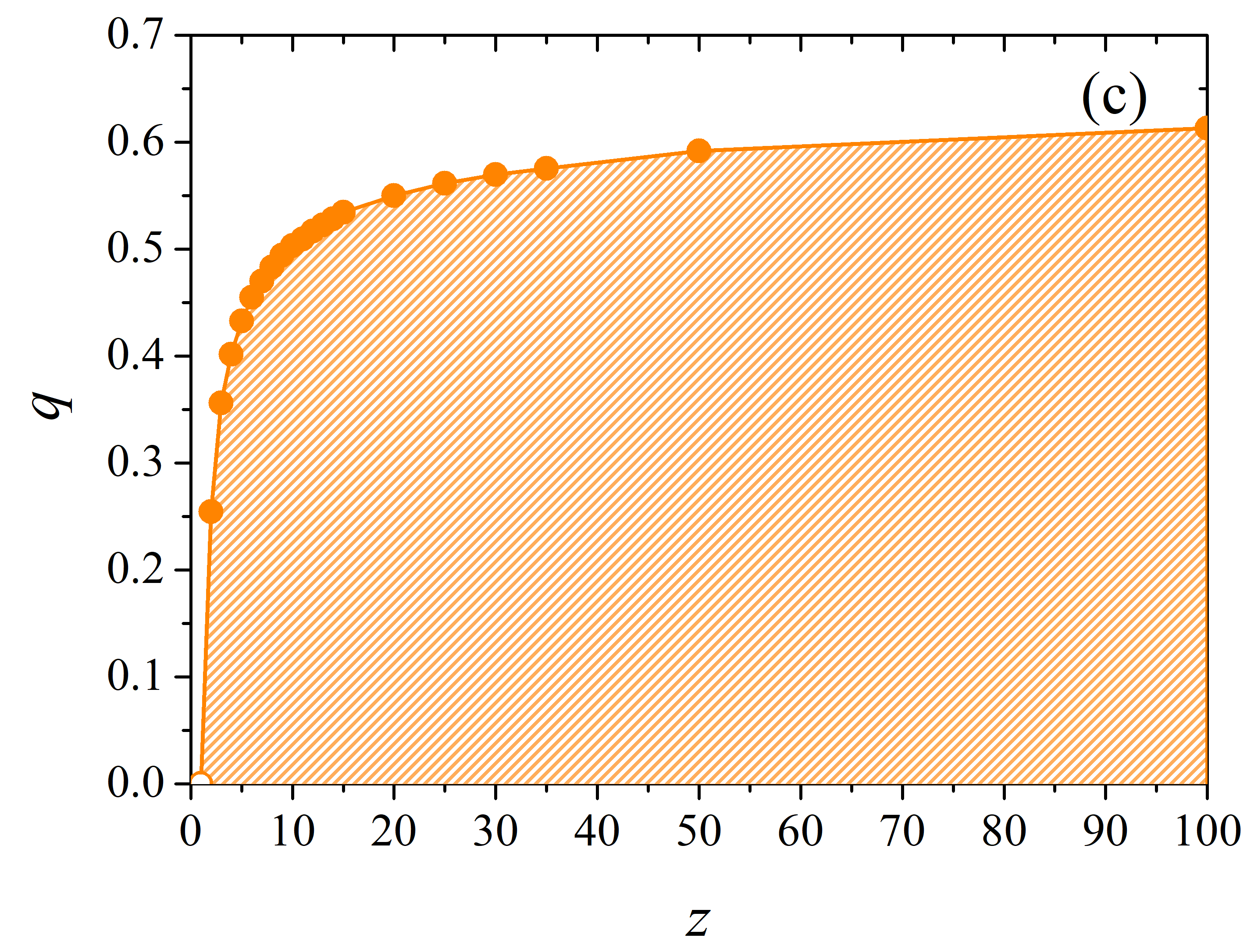}   
  \caption{(Color online) (a) The Binder's fourth-order cumulant $U(q, z, N)$ as a function of the noise parameter $q$ for the three-state majority-vote model in Barab\'asi-Albert networks with $z = 5$ for several values of the system sizes. From top to bottom we have $N = 20000, 15000, 10000, 5000, 2000$ and $1000$. In (b) we exhibit the details of the interception for different system sizes and a cubic fit for the data points in this region. Within the accuracy of the data, all curves intersect at $q_c = 0.4326(4)$. (c) Phase diagram of the three-state majority-vote model on Barab\'asi-Albert networks. The orange region denotes the phase where the system presents an order or a global majority opinion. The solid line is just a guide to the eye.
  }
  
    \label{binderpdiagram}
\end{figure*}

Figure \ref{binderpdiagram}(a) shows the dependence of the Binder's fourth-order cumulant on the noise parameter $q$ for $z = 5$ and different values of $N$. Note that the curves with different system sizes intercept when $0.43 < q < 0.44$. Figure \ref{binderpdiagram}(b) shows a magnification of the Binder cumulant data and a third order polynomial fit in the region near the curve interception, where the critical noise does not depend on the system size and is found to be $q_c = 0.4326(4)$. We calculate the cumulant $U(q, z, N)$ for other $z$ values and in Fig. \ref{binderpdiagram}(c) we show the phase diagram obtained for the three-state majority-vote model on Barab\'asi-Albert networks. The orange region denotes the ordered phase of the system, where one of the three opinions is the majority state of the system. In this result, the error bars are smaller than the thickness of the line.

\subsection*{The Unitary Relation and Scaling Results}

To obtain the critical exponents in complex networks, we propose that near the critical noise $q_c$ the correlation length $\xi$ scales with the actual volume of the system \cite{Botet1982, Hyunsuk2007} as 
\begin{equation}\label{xicomplex}
        \xi \sim N.
\end{equation}

\noindent Thus, the pseudocritical noise $q_c(N)$, the magnetization $M(q, z, N)$, the susceptibility $\chi(q, z, N)$, and the Binder cumulant $U(q, z, N)$ satisfy the finite-size scaling relations  
\begin{equation}\label{finiteq}
        q_{c}(N) = q_c + bN^{-1/\bar \nu},
\end{equation}
\begin{equation}\label{finitemag}
    M(q, z, N) = N^{-\beta/\bar \nu}\widetilde{M}(\epsilon N^{1/\bar \nu}),
\end{equation}
\begin{equation}\label{finitequi}
        \chi(q, z, N) = N^{\gamma/\bar \nu}\widetilde{\chi}(\epsilon N^{1/\bar \nu}),
\end{equation}
\begin{equation}\label{finitecumul}
        U(q, z, N) = \widetilde{U}(\epsilon N^{1/\bar \nu}),
\end{equation}
where $\epsilon = q - q_c$ is the distance to the critical noise, $b$
is a constant, and $\widetilde{M}$, $\widetilde{\chi}$, and
$\widetilde{U}$ are scaling functions that only depend on the scaled
variable $x = \epsilon N^{1/\bar\nu}$. For regular networks, we recall that $N = L^d$, where $d$ is the effective dimension of the network and $L$ is an effective linear size of the system. In this case, we obtain for the magnetization and for the magnetic susceptibility
\begin{equation}\label{finitemagd}
    M(q, z, N) = L^{-d\beta/\bar \nu}\widetilde{M}(\epsilon N^{1/\bar \nu}),
\end{equation}
\begin{equation}\label{finitequid}
        \chi(q, z, N) = L^{d\gamma/\bar \nu}\widetilde{\chi}(\epsilon N^{1/\bar \nu}).
\end{equation}
\noindent We use the notation $\bar \nu$ instead of $\nu$ since we changed the correlation length scaling relation from the usual linear scaling $\xi \sim L$ to the volumetric scaling $\xi \sim L^d$. In this case, the hyperscaling relation now reads $2\beta d/\bar \nu + \gamma d/\bar \nu = d$. Thus, we obtain 
\begin{equation}\label{hyperscaling}
        \frac{2\beta}{\bar \nu} + \frac{\gamma}{\bar \nu} = 1,
\end{equation}

\noindent regardless of the effective dimension $d$ of the network. This result allows us to remark that the hyperscaling relation cannot be used to estimate the dimension of these networks when using the volumetric scaling $\xi \sim L^d$, in contrast to the results of previous studies. \cite{Felipe2005, Lima2006, Lima2008, Melo2010} Nevertheless, the unitary relation (Eq. \ref{hyperscaling}) was verified in these works for random graphs and scale-free networks.  In this context, we rewrite the unitary relation by denoting a new exponent upsilon $\upsilon$, defined as 
\begin{equation}\label{upsilon}
        \upsilon \equiv \frac{2\beta}{\bar \nu} + \frac{\gamma}{\bar \nu},
\end{equation}

\noindent where we conjecture that $\upsilon = 1$ for any network under the condition of the volumetric scaling of Eq. (\ref{xicomplex}). In this work, we denote the equation $\upsilon = 1$  as the unitary relation for critical exponents. We validate the consistency of this result according to the comparison with the numerical findings for the critical exponents $\beta/\bar\nu$ and $\gamma/\bar\nu$ for regular and complex networks.

By calculating the logarithm of Eqs. (\ref{finiteq}), (\ref{finitemag}) and (\ref{finitequi}) at the critical point $q_c$, we obtain an explicit relation involving the critical exponents, the measured quantities and the system volume $N$

\begin{equation}\label{explifiniteq}
        \textrm{ln}[q_{c}(N) - q_c] \sim -\frac{1}{\bar \nu} \, \textrm{ln} \, N,
\end{equation}
\begin{equation}\label{explifinitemag}
    \textrm{ln}[M(q, z, N)] \sim -\frac{\beta}{\bar \nu} \, \textrm{ln} \, N,
\end{equation}
\begin{equation}\label{explifinitequi}
        \textrm{ln}[\chi(q, z, N)] \sim -\frac{\gamma}{\bar \nu} \, \textrm{ln} \, N,
\end{equation}

\noindent and we use the Equations (\ref{explifiniteq}), (\ref{explifinitemag}) and (\ref{explifinitequi}) to obtain the critical exponents of the system.

\begin{figure}
\centering
  \includegraphics[width=0.9\linewidth]{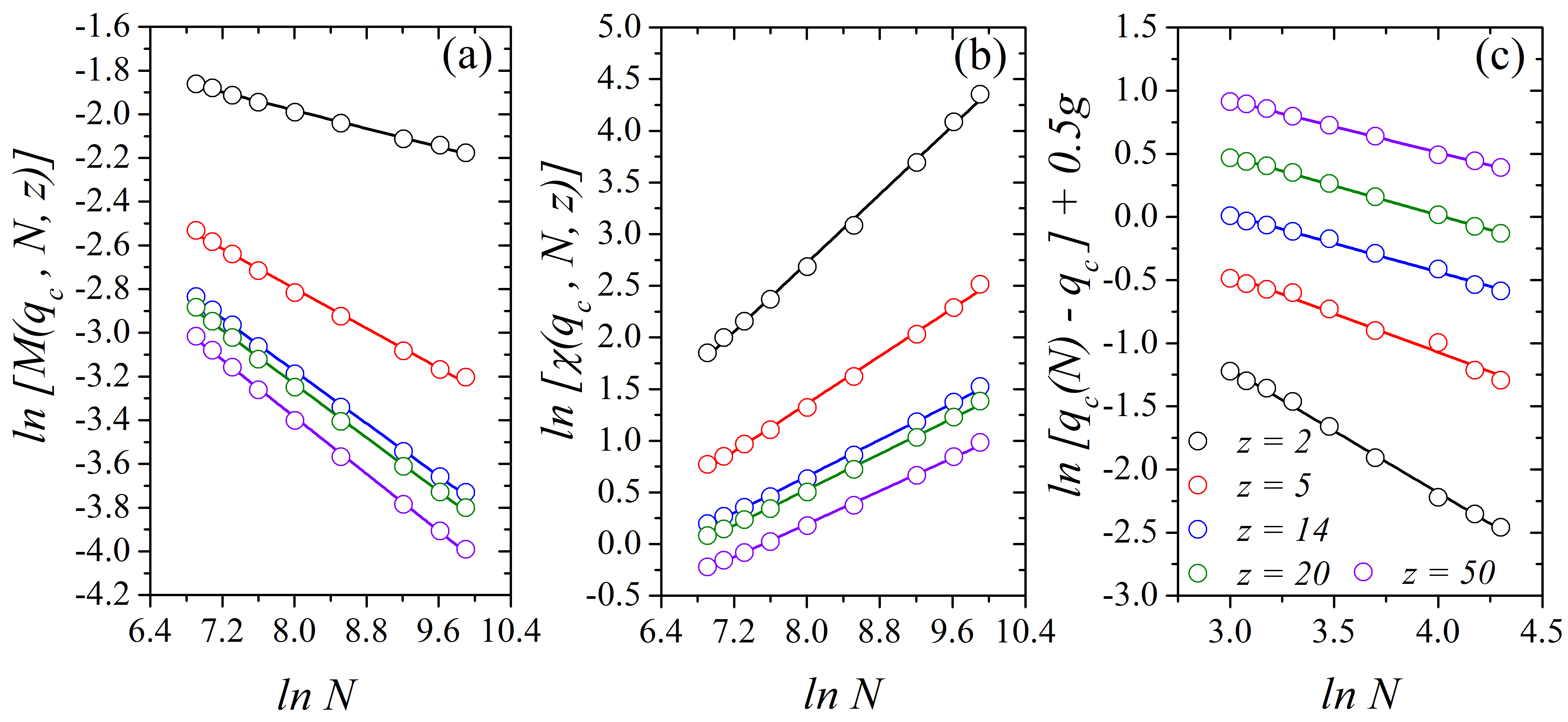}
  \caption{Plot at the critical point $q_c(z)$ for the logarithm of the (a) magnetization $\text{ln }M(q_c, z, N)$, (b) magnetic susceptibility $\text{ln }\chi(q_c, z, N)$, and $\text{ln }[q_c(N) - q_c]$ versus $\text{ln }N$ using $z = 2, 5, 14, 20$ and $50$, with $g = 1, 2, 3, 4$ and $5$. Here, $g$ is an integer used to better display the data. 
  }
  \label{loglines}
\end{figure}

Figure \ref{loglines} shows the logarithm of the (a) magnetization, of the (b) susceptibility and of the distance between the pseudocritical noise and the critical noise $[q_c(N) - q_c]$ versus the logarithm of the volume of the system $N$, where $q$ is set to be equal to $q_c(z)$. In this figure we show our results for $z = 2, 5, 14, 20$, and $50$, where the angular coefficient of the lines give us an estimation of the critical exponents $1/\bar\nu$, $\beta/\bar\nu$ and $\gamma/\bar\nu$. We find $\beta/\bar\nu = 0.102(4), 0.229(5), 0.300(3), 0.307(3)$ and $0.326(4)$, and $\gamma/\bar\nu = 0.83(1), 0.59(1), 0.44(1), 0.43(1)$ and $0.40(1)$, and $1/\bar\nu = 1.01(2), 0.61(3), 0.45(1), 0.47(1)$ and $0.41(1)$, for $z = 2, 5, 14, 20$, and $50$, respectively. 

Table~\ref{texp} provides the critical noise, the critical exponents and the unitary relation values for each growth parameter investigated in the model. Note that the critical exponents for the magnetization (susceptibility) is a decreasing (an increasing) function of the growth parameter $z$. The critical noise $q_c$ increases with $z$, while the critical exponent $1/\bar\nu$ decrease with $z$. For all values in the table, we obtain $\upsilon \sim 1$ as expected.

\begin{table}[ht]
\centering
\begin{tabular}{|l|l|l|l|l|l|}
\hline
$z$ & $q_c$ & $\beta/\bar \nu$ & $\gamma/\bar \nu$ & $1/\bar \nu$ & $\upsilon$ \\
\hline
$2$ & $0.2549(3)$ & $0.102(4)$ & $0.83(1)$ & $1.01(2)$ & $1.04(3)$\\
\hline
$3$ & $0.3561(5)$ & $0.141(4)$ & $0.75(1)$ & $0.82(3)$ & $1.03(2)$\\
\hline
$4$ & $0.4015(5)$ & $0.197(5)$ & $0.64(1)$ & $0.65(2)$ & $1.03(2)$\\
\hline
$5$ & $0.4326(4)$ & $0.229(5)$ & $0.59(1)$ & $0.61(3)$ & $1.03(2)$\\
\hline
$6$ & $0.4699(6)$ & $0.249(5)$ & $0.54(1)$ & $0.53(1)$ & $1.04(2)$\\
\hline
$7$ & $0.4699(6)$ & $0.244(5)$ & $0.54(1)$ & $0.53(1)$ & $1.03(2)$\\
\hline
$8$ & $0.4832(5)$ & $0.261(4)$ & $0.51(1)$ & $0.51(1)$ & $1.04(2)$\\
\hline
$10$ & $0.5031(9)$ & $0.299(2)$ & $0.45(1)$ & $0.48(1)$ & $1.04(1)$ \\
\hline
$14$ & $0.5282(1)$ & $0.300(3)$ & $0.44(1)$ & $0.45(1)$ & $1.04(1)$ \\
\hline
$20$ & $0.5500(5)$ & $0.307(3)$ & $0.43(1)$ & $0.47(1)$ & $1.05(1)$\\
\hline
$25$ & $0.5617(1)$ & $0.309(4)$ & $0.43(1)$ & $0.44(1)$ & $1.05(2)$\\
\hline
$50$ & $0.5918(1)$ & $0.326(4)$ & $0.40(1)$ & $0.41(1)$ & $1.05(1)$\\
\hline

\end{tabular}
\caption{\label{texp}The critical noise $q_c$, the critical exponents $\beta/\bar\nu$, $\gamma/\bar\nu$ and $1/\bar\nu$, and the unitary relation $\upsilon$, for the three-state majority-vote model on Barab\'asi-Albert networks with growth parameter $z$.}
\end{table}

\begin{figure*}[t]
  \centering
    \includegraphics[width=0.6\linewidth]{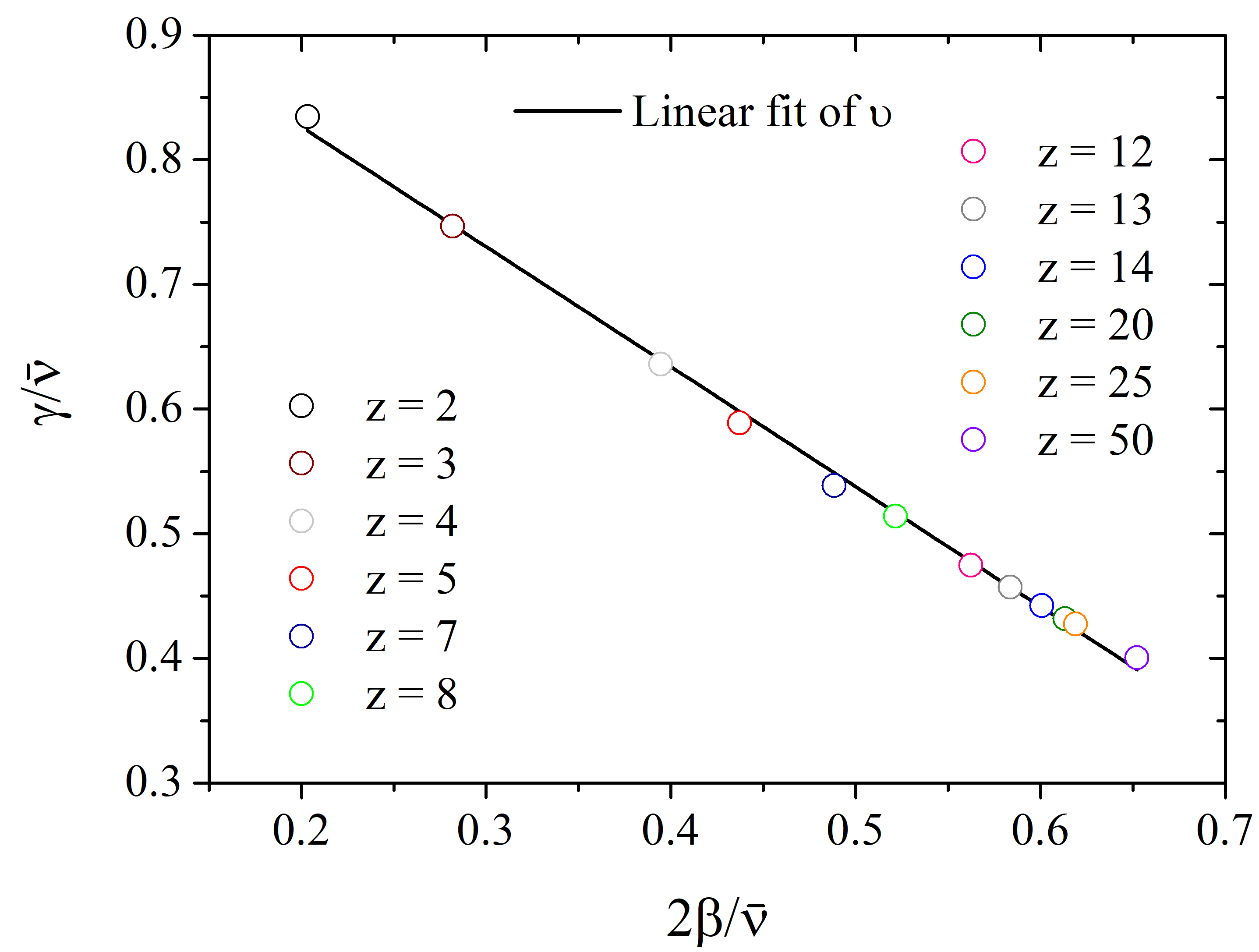}  
  \caption{(Color online) Plot of the unitary line $y = -0.96(1)x+1.02(1)$ estimated by the linear fit of the relation between the critical exponents $\beta/\bar\nu$ and $\gamma/\bar\nu$ for several values of the growth parameter $z$.
    }
    \label{unitaryrelation}
\end{figure*}

By using the critical exponents, we obtain the unitary line of the model showed in Figure \ref{unitaryrelation}. Here, we plot the values of the critical exponents $\gamma/\bar\nu$ versus $2\beta/\bar\nu$. The linear fit of the data yield $y = -0.96(1)x+1.02(1)$, with an averaged unitary exponent $\upsilon = 1.02(1)$ for the three-state majority-vote model on Barab\'asi-Albert networks. We conjecture that this line is universal, regardless of the geometric structure of the network used in the model. Thus, one can use the volumetric scaling $\xi \sim N$ with the unitary relation (\ref{upsilon}), and the Equations (\ref{explifiniteq}), (\ref{explifinitemag}) and (\ref{explifinitequi}) to obtain the critical exponents and the unitary line for a spin model being investigated, with or without a linear system size length clearly defined. 

\begin{figure*}[t]
  \centering
    \includegraphics[width=0.9\linewidth]{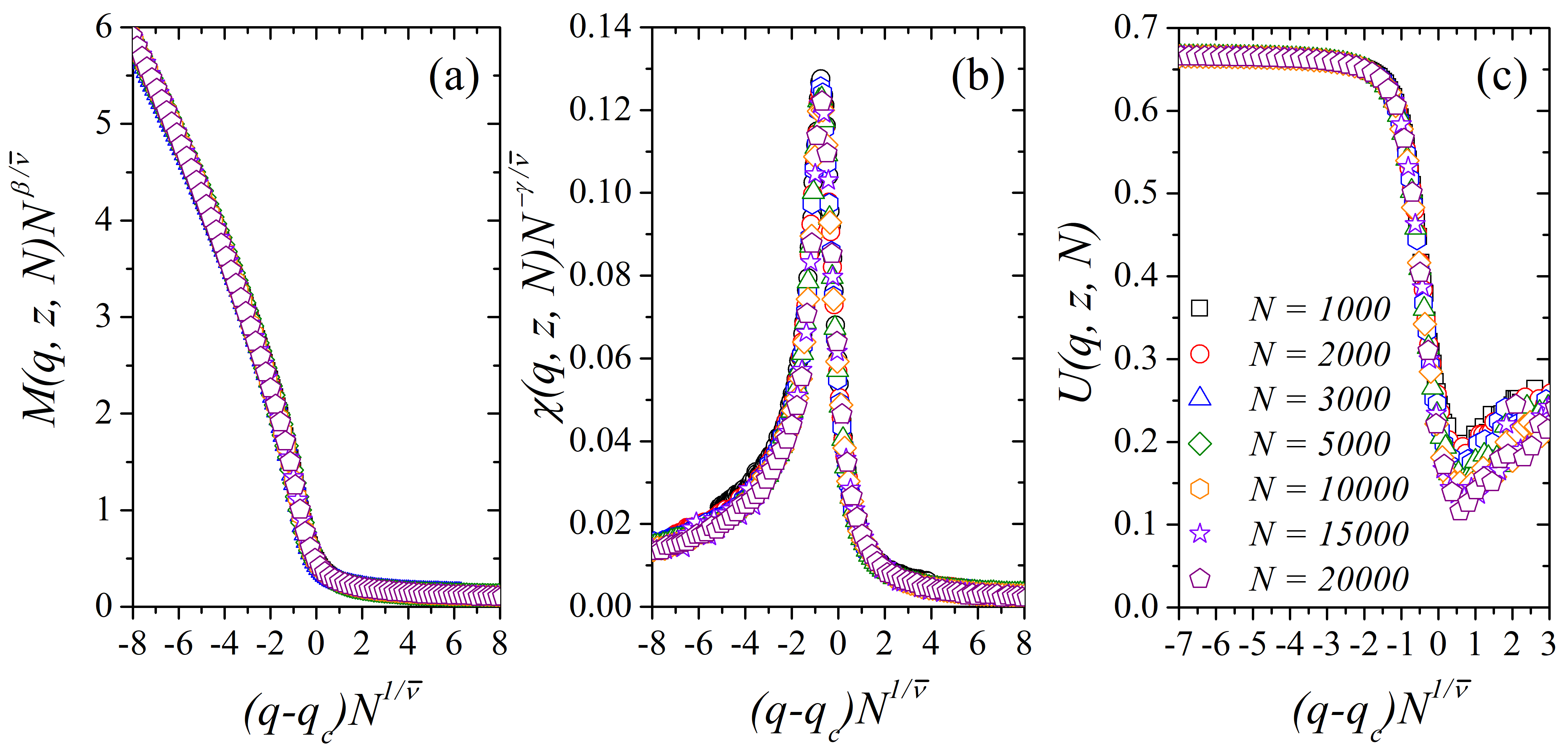}  
  \caption{(Color online) Data collapse for (a) the magnetization $M(q, z, N)$, (b) the magnetic susceptibility $\chi(q, z, N)$ and (c) the Binder's fourth-order cumulant $U(q, z, N)$ for $N = 1000, 2000, 3000, 5000, 10000, 15000$ and $20000$ with $z = 14$. Here, we used $\beta/\bar\nu = 0.3$, $\gamma/\bar\nu = 0.44$ and $1/\bar\nu = 0.45$.  
  }
    \label{collapse14color}
\end{figure*}

Figure~\ref{collapse14color} shows the plot of the rescaled (a) magnetization $M(q, z, N)N^{\beta/\bar\nu}$, (b) susceptibility $\chi(q, z, N)L^{-\gamma/\bar\nu}$, and (c) Binder cumulant $U(q, z, N)$ versus rescaled noise parameter $(q - q_c)N^{1/\bar\nu}$ for the growth parameter $z = 14$. Here, we used $\beta/\bar\nu = 0.300$, $\gamma/\bar\nu = 0.44$ and $1/\bar\nu = 0.45$ with $q_c = 0.5282$. Other growth parameter values $z$ exhibit the same features and the same qualitative results for the data collapse of the magnetization, susceptibility, and Binder cumulant. From our simulation results and analysis, we conclude that the three-state majority-vote model defined on Barab\'asi-Albert networks and on the Erd\H{o}s–R\'enyi random graphs belong to different universality classes when the volumetric scaling $\xi \sim N$ is used. \cite{Melo2010}

\subsection*{Unitary Relation on Regular Networks}

To confirm the validity of our statements, we performed Monte Carlo simulations for the majority-vote model with two and three states on regular square lattices and on cubic networks. We also obtain the critical noise and the critical exponents for the three-state majority-vote model in cubic networks.

\begin{figure*}[t]
  \centering
    \includegraphics[width=0.9\linewidth]{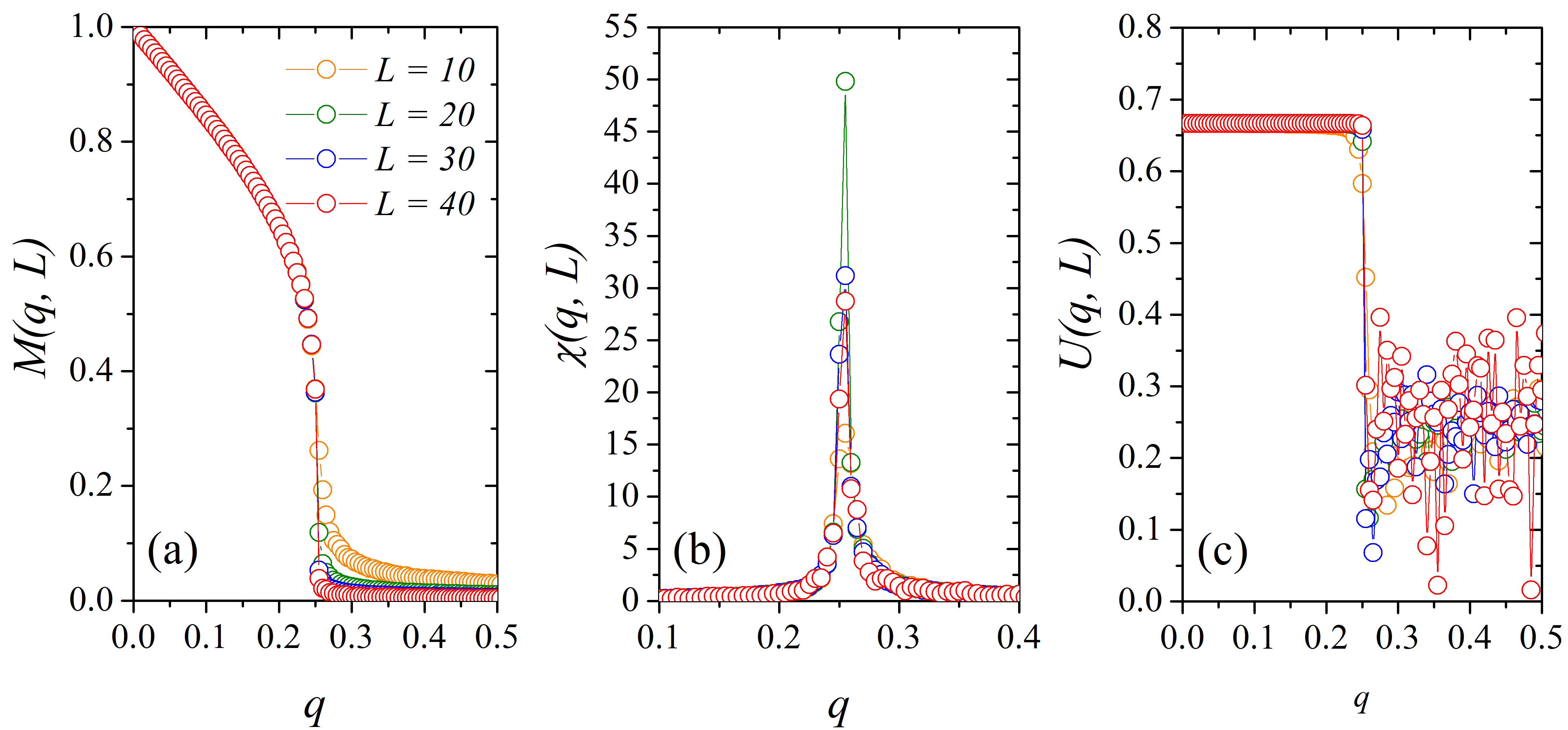}  
  \caption{(Color online) (a) Magnetization $M(q, L)$, (b) magnetic susceptibility $\chi(q, L)$ and (c) Binder cumulant $U(q, L)$ for the three-state majority-vote model on cubic networks with $L = 10, 20, 30$ and $40$.  
  }
    \label{overview3st3d}
\end{figure*}

For the three-state majority-vote model on cubic networks with $L = 10, 20, 30$ and $40$, with periodic boundary conditions. We use $3 \times 10^5$ MCS to averages our quantities, $10^5$ MCS as thermalization time and we generated $100$ independent random samples in order to calculate the configurational averages. Figure \ref{overview3st3d} shows the our results for the (a) magnetization $M(q, L)$, the (b) susceptibility $\chi(q, L)$ and the (c) Binder cumulant $U(q, L)$ versus the noise parameter $q$, where $N = L^3$. We observe some familiar results such as $M(q, L) \to 0$ for $q > q_c$, with $L \to \infty$, and $\chi(q, L)$ that also exhibits a sharper peak as we increase $L$. From Binder parameter, we find the critical noise of the model $q_c = 0.25230(2)$.

By performing numerical simulations for the majority-vote model with two and three states on regular square lattices and on cubic networks we build the Table \ref{tregular} with the critical exponents for the magnetization and susceptibility. We also obtain unitary relation $\upsilon$ and effective dimension $d$ obtained for each model with volumetric ($\xi \sim N$) and linear ($\xi \sim L$) scalings, respectively. We conclude that the critical exponents with linear and volumetric scalings relate by $\beta/\nu = d \beta/\bar\nu$ and $\gamma/\nu = d \gamma/\bar\nu$, as expected by recalling that $\xi \sim L^d$.

\begin{table}[ht]
\centering
\begin{tabular}{|l|l|l|l|l|l|l|l|l|}
\hline
Network & States ($Q$) & $q_c$ & $\beta/\bar \nu$ & $\gamma/\bar \nu$ & $\upsilon$ & $\beta/ \nu$ & $\gamma/ \nu$ & $d$\\
\hline
$Square$ & $3$ & $0.118(1)$ & $0.067(1)$ & $0.90(1)$ & $1.03(1)$ & $0.134(1)$ & $1.80(1)$ & $2.07(1)$\\
\hline
$Square$ & $2$ & $0.075(1)$ & $0.062(1)$ & $0.87(1)$ & $0.99(1)$ & $0.124(1)$ & $1.74(1)$ & $1.99(1)$\\
\hline
$Cubic$ & $3$ & $0.2523(1)$ & $0.197(5)$ & $0.64(1)$ & $1.03(2)$ & $0.33(1)$ & $2.41(1)$ & $3.07(3)$\\
\hline
$Cubic$ & $2$ & $0.1761(3)$ & $0.154(2)$ & $0.70(1)$ & $1.01(1)$ & $0.461(7)$ & $2.11(1)$ & $3.03(2)$\\
\hline

\end{tabular}
\caption{\label{tregular} The critical noise $q_c$, the critical exponents volumetrically rescaled for $\xi \sim L^d$, the unitary relation $\upsilon$, the regular critical exponents $\beta/\nu$ and $\gamma/\nu$, and the effective dimension $d$ when $\xi \sim L$ for the majority-vote model with two and three states on regular networks.}
\end{table}

\begin{figure*}[t]
  \centering
    \includegraphics[width=0.49\linewidth]{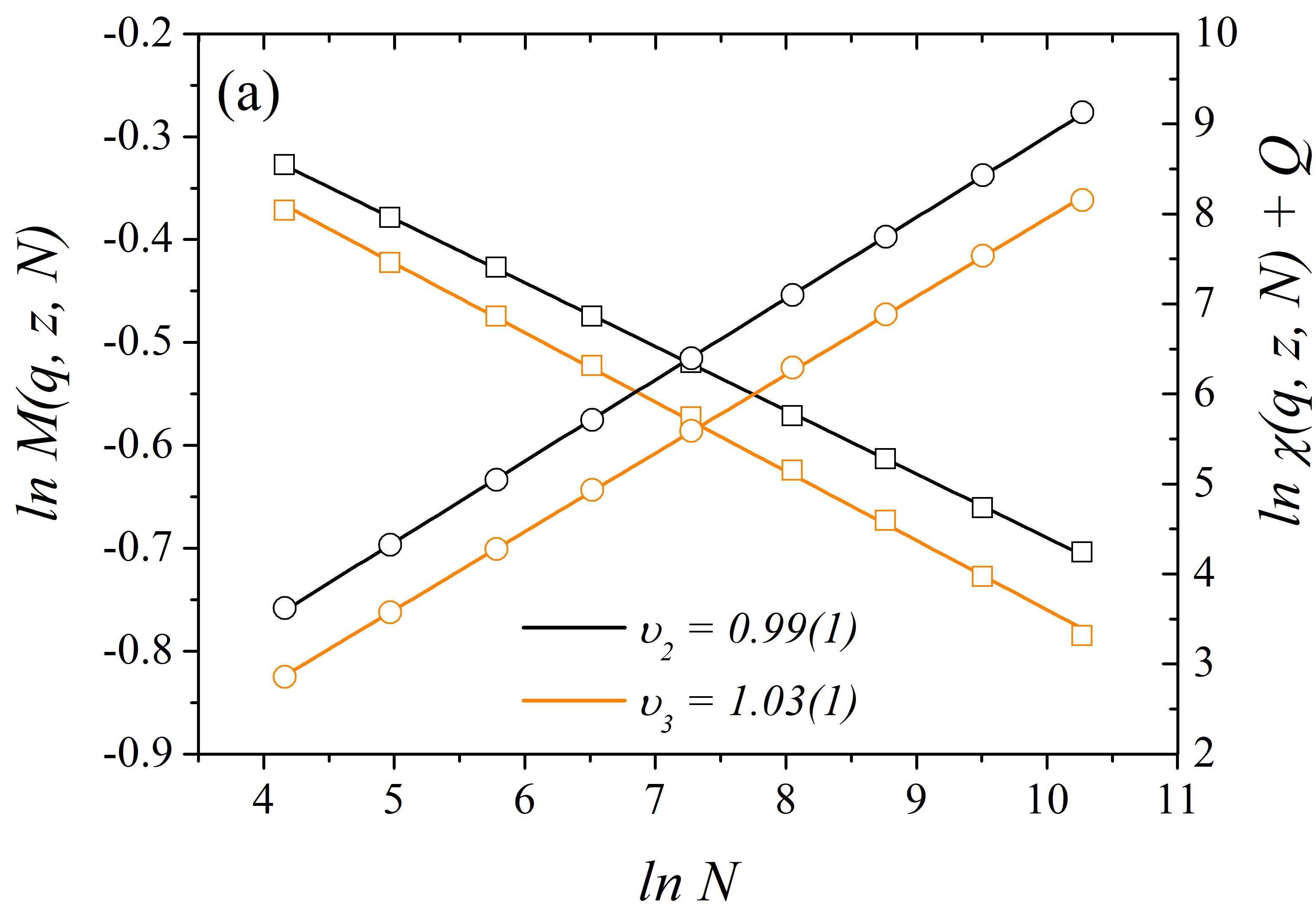}
    \includegraphics[width=0.447\linewidth]{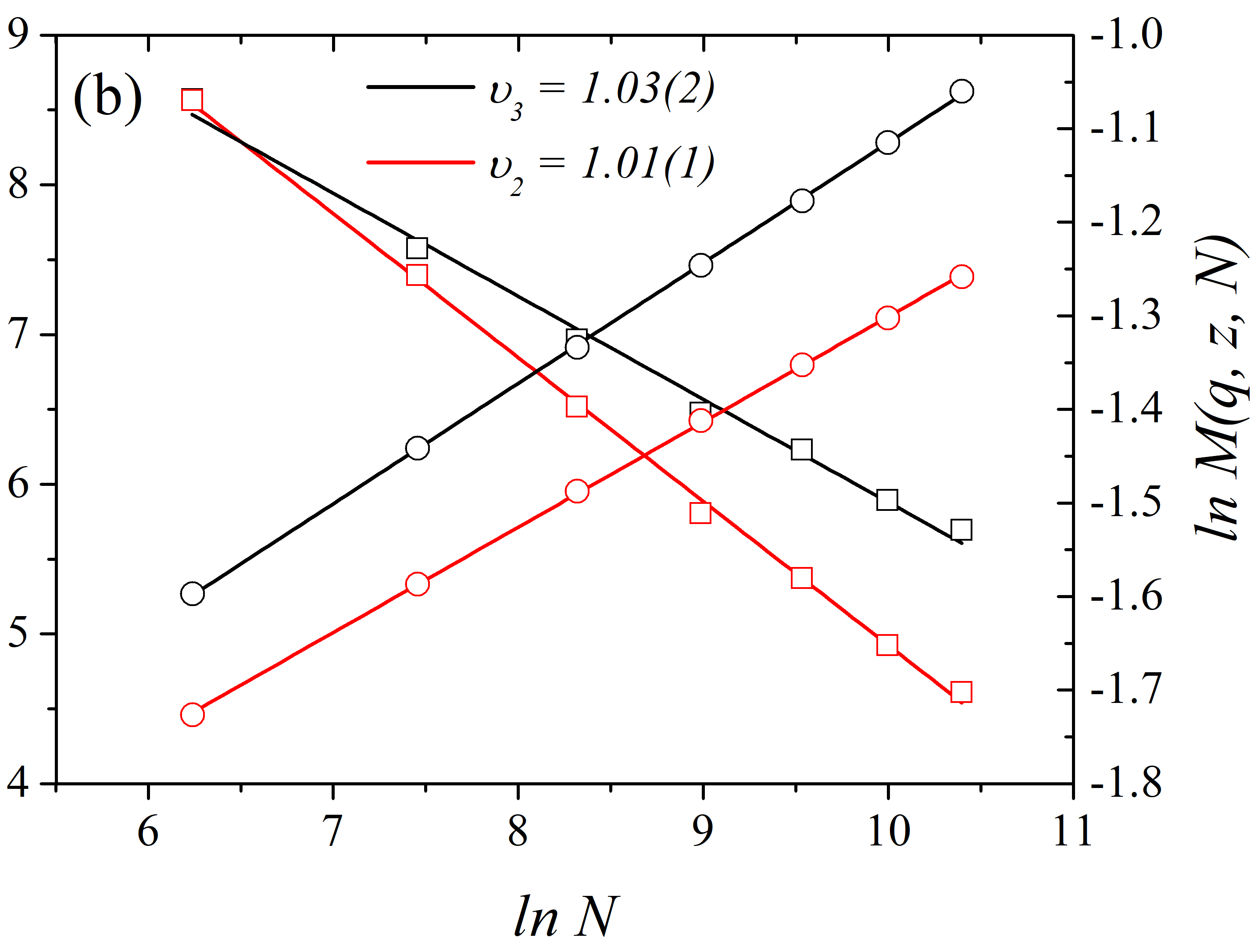}
  \caption{(Color online) Logarithm of the magnetization (squares) and of the susceptibility (circles) for the majority-vote model with $Q = 2$ and $Q = 3$ states as a function of the logarithm of the system sizes $N$ in for (a) square lattice and (b) cubic networks. We obtain for regular square lattices $\beta/\bar\nu = 0.067(1)$, $\gamma/\bar\nu = 0.90(1)$ and $\upsilon_3 = 1.03(1)$ for $Q = 3$, and $\beta/\bar\nu = 0.062(1)$, $\gamma/\bar\nu = 0.87(1)$ and $\upsilon_2 = 0.99(1)$ for $Q = 2$. For a cubic network (b), we obtain $\beta/\bar\nu = 0.107(5)$, $\gamma/\bar\nu = 0.81(1)$ and $\upsilon_3 = 1.02(2)$ for $Q = 3$, and $\beta/\bar\nu = 0.154(2)$, $\gamma/\bar\nu = 0.70(1)$ and $\upsilon_2 = 1.01(1)$ for $Q = 2$.
    }
    \label{upsilonregular}
\end{figure*}

In Figure \ref{upsilonregular} we plot the logarithm of the magnetization and magnetic susceptibility show the critical exponents obtained for the majority-vote model with two and three states on square lattice and cubic networks with the volumetric scaling. Our results confirms that the unitary relation holds for this model on these networks, and it points that the effective dimension obtained by previous works with the majority-vote model on random graphs and on Barab\'asi-Albert networks might be not equal to unity. \cite{Felipe2005, Lima2006, Lima2008, Melo2010}

\section*{Conclusion and Final Remarks}

 We have obtained the phase diagram and critical exponents of the three-state majority-vote model with noise on Barab\'asi-Albert networks. We verified that the second-order phase transition occurs in for networks with growth parameter $z > 1$, where the critical exponents $\gamma/\bar \nu$ and $1/\bar \nu$ decrease with $z$, while $\beta/\bar \nu$ increases. We also find that the critical noise $q_c$ is an increasing function of the growth parameter $z$. By assuming that near criticality the correlation length $\xi$ scales with the actual volume of the system $\xi \sim N$, we found that the hyperscaling relation leads to unity, regardless the effective dimension of the network of interactions. Nevertheless, using Monte Carlo simulations we verified the unitary relation $\upsilon = 1$ for all values of the growth parameter $z$ investigated. The unitary relation was also confirmed for the majority-vote model with two and three states on regular square lattices and cubic networks, with well defined effective dimensions. 
  
 We remark that obtaining the effective dimension for some complex networks - such as the random graphs and the Barab\'asi-Albert networks - using the relations between critical exponents remains as a task to perform, although other methods and scalings have been successfully developed. \cite{Daqing2011, Berche2012} Our results, may suggest and allow future research using other finite-size scaling relations, such as some power-law with a characteristic length instead. We also suggest using the unitary relation and the unitary line as tests and confirmation for the criticality of systems with complex interactions of unknown effective dimension or clearly defined linear size.
 
\bibliography{MVM3_BA-refs}

\section*{Acknowledgements}

The authors acknowledge financial support from UPE (PFA2018, PIAEXT2018)
and the funding agencies FACEPE (APQ-0565-1.05/14, APQ-­0707­-1.05/14), CAPES and CNPq. This work has been supported in part by the National Natural Science Foundation of China (61603011), Beijing Social Science Foundation (16JDGLC005), and International Postdoctoral Exchange Fellowship Program (20170016). The Boston University Center for Polymer Studies is supported by NSF Grants PHY-1505000, CMMI-1125290, and CHE-1213217, by DTRA Grant HDTRA1-14-1-0017, and by DOE Contract DE-AC07-05Id14517. 

\section*{Author contributions statement}

In this work, all authors wrote the main manuscript text, A.V. also performed the simulations and prepared the figures. All authors reviewed the manuscript. 

\section*{Additional information}
\textbf{Competing interests:} The authors declare no competing interests.

\end{document}